\documentclass{article}
\begin{document}

 \title{Magnetic translation and Berry's phase factor through adiabatically rotating a magnetic field \thanks
 {An earlier version published in Physics Letters A ${\bf 275}$ (2000) 473-480.}}

 \author{J. Chee \thanks{Current address: Physics Department, University of Lethbridge, Lethbridge, AB T1K 3M4, Canada} 
 \\\\email: jameschee@email.com}
 \date{April 9, 2002}
 \maketitle
 \begin{abstract}
 For a spin subjected to an adiabatically changing magnetic field, the "solid angle result" as embodied by a rotation
 operator is the only nontrivial factor in the evolution operator. For a charged particle, the infinite
 degeneracy of the energy levels calls for a rigorous investigation. We find that in this case, it is the product of 
 the rotation operator
and a magnetic translation operator that enters into the evolution operator and determines the 
 geometric phase. This result agrees with the fact that the instantaneous hamiltonian is invariant under magnetic 
 translation as well
as rotation. Experimental verification of the result is proposed.
 \end{abstract}


\section{Introduction}

The quantum evolution of a magnetic dipole in an adiabatically changing magnetic field \cite{berry} has provided 
the classic example of the Berry phase. What happens if a charged particle moves in an adiabatically changing magnetic field? 
Is the solution similar to the magnetic dipole case as is often assumed in the literature? We point out in this paper that 
for the charged particle case, we have a beautiful answer to the above questions because of the magnetic translation symmetry 
of the instantaneous Hamiltonian. 
  
In the proof of the quantum adiabatic theorem as presented in Messiah's book \cite{messiah}, the quantum evolution
operator is constructed as the product of a path-dependent factor and a dynamical factor. With the discovery of the
Berry phase phenomenon \cite{berry}, it is clear that the path-dependent factor in the evolution operator has
nontrivial consequences; namely, after a cyclic change of the parameters, this factor is in general
not equal to the identity operator but rather should recover a Berry phase factor for an adiabatic eigenstate of the
Hamiltonian. When a finite fold degeneracy is involved, there is the non-Abelian generalization of 
the Berry phase
concept due to Wilczek and Zee \cite{wil}. It remains to be rigorously investigated as to why and how the 
adiabatic theorem applies when an infinite degeneracy is involved. It can be seen that in all of the above situations,
the factorisation of the evolution operator captures the essence of the quantum adiabatic theorem (when it applies) 
incorporating the Berry phase phenomenon.

Now we come to the problem of a charged particle moving in a slowly rotating magnetic field. We 
shall choose a harmonic oscillator potential in the (changing) direction of the magnetic field. The
purpose is to get rid of the unbounded motion of the particle in that direction. The infinite degeneracy is still retained 
beause of the infinite degeneracy of the Landau levels. The underlying symmetry that is responsible for this degeneracy is 
just the magnetic translation, as was studied in Refs \cite{brown} and \cite{Zak}. When there is no potential in the 
magnetic
field direction, we give a discussion of the problem at the end of the paper. Whether there is any interesting 
result in this case is a question that may deserve further discussion.

The vector potential for a uniform but slowly rotating magnetic field can be written as
\begin{equation}\label{1}
{\bf A}(\epsilon t)={1\over2}B{\bf{n}}(\epsilon t)\times\bf{x},
\end{equation}
where $B$ is constant and ${\bf{n}}(\epsilon t)$ is a unit vector representing the direction of the magnetic field.
Following Messiah \cite{messiah}, the time dependence of the parameters and therefore the Hamiltonian is through a
slowness parameter $\epsilon =\frac{1}{T}$, where $T$ is the duration of the adiabatic process. An adiabatic process
means that $T$ is much larger than all the other time scales involved. We have the following Hamiltonian:
\begin{equation}\label{2}
  H=\frac{1}{2m}\big({\bf P}-{q\over c}{\bf A}(\epsilon t)\big)^2+
  {1\over 2}m\omega^2\big({\bf n}(\epsilon t)\cdot{\bf x}-a\big)^2,
\end{equation}
where $a$ represents the equilibrium position of the oscillator
potential along the direction of ${\bf n}(\epsilon t)$.

A solenoid slowly rotating about a fixed point on the symmetry axis generates a magnetic field inside the solenoid
that is described by the vector potential (1). Equation (1) singles out a unique point in space, ${\bf{x}}=\bf{0}$,
where the induced electric field $-\frac{1}{c}\frac{\partial\bf A}{\partial t}$ vanishes. Such a point corresponds
to the fixed point about which the solenoid rotates, and ${\bf n}(\epsilon t)$ is along the direction of the
symmetry axis. There are also higher order induced electro-magnetic field due to the slow change of
$-\frac{1}{c}\frac{\partial\bf A}{\partial t}$ and so on that are not included by (2). However, such effects are of
the order of $\frac{1}{T^2}$ and therefore do not accumulate in the adiabatic process with $t\in[0, T]$. It should
also be noticed that potential (1) describes the region inside the solenoid only.

We shall obtain the quantum evolution operator corresponding to (2) in the adiabatic limit of $T\rightarrow\infty$
as the product of a path-dependent geometrical operator and a usual dynamical operator. The path considered here is
the path of $\bf n$ on the two-dimensional sphere.

Observe that a Berry phase factor is often associated with an adiabatic eigenstate of the Hamiltonian. But it is
equivalent to focus on the geometrical operator because Berry's phase factor is obtained at the end of the cycle by
letting the geometrical operator act on an eigenstate of the initial Hamiltonian. The non-Abelian generalization
\cite{wil} of the Berry phase factor can also be recovered from matrix elements of the geometrical operator among
the degenerate eigenstates. Therefore, the Berry phase factor can be defined as the geometrical operator when $\bf
n$ returns back to ${\bf n}(0)$. Of course, the geometrical operator is more general and applies for all $t\in [0,
T]$ even when ${\bf n}(1)$ is not equal to ${\bf n}(0)$.

We will find that, in the context of model (2), a magnetic translation operator \cite{brown,Zak,Kohm} plays a
natural role in addition to the rotation operator well known from Berry's spin example \cite{berry}. It will be
shown that the path-dependent factor in the quantum evolution operator is exactly the product of the rotation
operator and a path-ordered magnetic translation operator.

Our method is to solve the Heisenberg equations first. We will find an operator $U$ (in a factorised form) such that
any physical observable $O(t)$ is given by $O(t)$=$U^\dag(t)O(0)U(t)$. This $U$ is the evolution operator up to a
phase factor. This is because any $e^{i\alpha(t)} U$ where $\alpha(t)$ is a real number also solves the Heisenberg
equations. By Schur's lemma, this is also the only ambiguity. To determine the evolution operator uniquely, we must
make use of the Schr\"{o}dinger picture. The quantum evolution operator is finally determined in a factorised form
in equation (50).

\section{Solving Heisenberg Equations Using an Operator}
In this section we solve the Heisenberg equations for $\bf x$ and $\bf P$ in the adiabatic limit by using an
operator $U(t)= R(\epsilon t)M(\epsilon t)D(t)$, such that $x_{i}(t)=U^{\dag}(t)x_{i}(0)U(t)$,
$P_{i}(t)=U^{\dag}(t)P_{i}(0)U(t)$. This $U(t)$ assumes a factorised form, where $R(\epsilon t)$ is a rotation
operator, $M(\epsilon t)$ is a magnetic translation operator, and $D(t)=e^{-iH(0)t/\hbar}$.

The Heisenberg equations $\dot{x}_{i}=\frac{1}{i\hbar}[x_{i}, H]$ and $\dot{P}_{i} = \frac{1}{i\hbar}[P_{i}, H]$,
upon using the commutation relations $[x_{i}, P_{j}] = i\hbar\delta_{ij}$ and $[x_{i}, x_{j}] = [P_{i}, P_{j}] = 0$,
result in the following:
\begin{equation}
\dot{\bf x}=\frac{1}{m}{\bf P}-\frac{\omega_c}{2}{\bf n}\times\bf
x,
\end{equation}

\begin{equation}
\dot{\bf P}=-m\omega^2{\bf n}({\bf n}\cdot{\bf
x}-a)-\frac{\omega_c}{2}{\bf n}\times({\bf
P}-\frac{m\omega_c}{2}{\bf n}\times{\bf x}),
\end{equation}
where $\omega_c=\frac{qB}{mc}$.

We shall solve the Heisenberg equations in a rotating frame specified by the unit vector ${\bf
e}_3={\bf{n}}(\epsilon t)$ and two other unit vectors ${\bf e}_1(\epsilon t)$ and ${\bf e}_2(\epsilon t)$ that are
determined by certain requirements (equation (7) and initial conditions). By $\rm (i)$ relating components of
operators in this frame and the corresponding ones in the stationary frame $\{{\bf e}_{i}(0), i=1,2,3\}$ using a
rotation operator $R$ and $\rm (ii)$ solving for $\bf x$ and $\bf P$ in the rotating frame $\{{\bf e}_{i}(\epsilon
t), i=1,2,3\}$; $U$ is constructed. The stationary frame is just the reference frame with respect to which the
Hamiltonian (2) is written. The $x_{i}$ and $P_{i}$ mentioned so far are components of $\bf x$ and $\bf P$ with
respect to this frame. In the following, repeated Latin indices are summed from 1 to 3 and repeated Greek indices
are summed from 1 to 2.

Let

\begin{equation}
  {\bf x}(t) = \tilde{x}_{i}(t){\bf e}_{i}(\epsilon t)=x_{i}(t){\bf
  e}_{i}(0),
\end{equation}
\begin{equation}
{\bf P}(t) = \tilde{P}_{i}(t){\bf e}_{i}(\epsilon t)=P_{i}(t){\bf
e}_{i}(0),
\end{equation}
where ${\bf e}_{i}(\epsilon t), i=1, 2, 3,$ are determined by the
initial values ${\bf e}_{i}(0)$ and the following equation

\begin{equation}
  \dot{\bf e}_{i} = ({\bf n}\times\dot{\bf n})\times{\bf e}_{i}.
\end{equation}
We demand that ${\bf e}_{i}(0)$ are unit vectors and ${\bf
e}_{3}(0)={\bf n}(0)={\bf e}_{1}(0)\times{\bf e}_{2}(0)$. The
meaning of equation (7) is that ${\bf e}_i$ rotates with the
instantaneous angular velocity $\bf n\times\dot{\bf n}$ that is
perpendicular to $\bf n$. From equation (7) and the initial
condition, we get

\begin{equation}
{\bf e}_{i}(0)\cdot\dot{\bf e}_{j}(\epsilon t)=\epsilon_{imk}({\bf
n}(\epsilon t)\times\dot{\bf n}(\epsilon t))\cdot{\bf
e}_{m}(0)\big({\bf e}_{k}(0)\cdot{\bf e}_{j}(\epsilon t)\big).
\end{equation}
Now consider the matrix $E$ whose matrix elements are:
\begin{equation}
E_{ij}(\epsilon t)={\bf e}_{i}(0)\cdot{\bf e}_{j}(\epsilon t).
\end{equation}
By equations (8) and (9),

\begin{equation}\label{10}
E = P\exp\int\limits_{{\bf n}(0)}^{{\bf n}(\epsilon t)} \ J_{m}({\bf n}\times d{\bf n}) \cdot {\bf e}_{m}(0),
\end{equation}
where $P\exp$ means ``path-ordered exponential" and where $J_{m}, m=1,2,3$ are $3 \times 3$ matrices whose matrix
elements are given by

\begin{equation}\label{11}
(J_{m})_{ik} = - \epsilon_{mik}.
\end{equation}
The matrix $E$ determines ${\bf e}_{i}$ by the relation

\begin{equation}\label{12}
{\bf e}_{i}(\epsilon t) = {\bf e}_{k}(0)E_{ki}(\epsilon t).
\end{equation}
From equation (7) and initial conditions, it follows that ${\bf e}_{3}={\bf n}$ for all $t$ and that ${\bf
e}_{i}\cdot{\bf e}_{j}=\delta_{ij}$. The magnetic field $\bf B$ becomes a constant in this frame. Moreover, ${\bf
e}_{1}$ and ${\bf e}_{2}$ can be seen as tangent vectors to the two-dimensional sphere and satisfy (from Eq.(7)):

\begin{equation}\label{13}
\dot{\bf e}_{\mu}\cdot{\bf e}_{\nu}=0; \, \mu, \nu=1,2.
\end{equation}
Therefore ${\bf e}_{1}$ and ${\bf e}_{2}$ have the geometrical meaning of being parallel transported on the 2-sphere
along the path of $\bf n$.

If $\bf x$ and $\bf P$ are found in this frame, the components of $\bf x$ and $\bf P$ with respect to the constant
frame $\{{\bf e}_{i}(0)\}$ are known through equations (5) and (6). However this is not enough, because we want to
solve for $x_{i}$ and $P_{i}$ using an operator. So we need to find the operator that relates the rest frame
operators ($x_i$ and $P_i$) and the moving frame operators ($\tilde{x}_i$ and $\tilde{P}_i$) which is done as
follows.

Suppose $\tilde{U}(t)$ expresses the evolution of $\bf x$ and $\bf P$ in the frame ${\bf e}_{i}(\epsilon t)$, such
that

\begin{equation}\label{14}
 \tilde{x}_{i}(t) =\tilde{U}^{\dag}(t)x _{i}(0) \tilde{U}(t),
\end{equation}\label{15}
\begin{equation}
 \tilde{P_{i}}(t) =\tilde{U}^{\dag}(t)P _{i}(0) \tilde{U}(t).
\end{equation}
Then by equation (5),
\begin{equation}\label{16}
x_{i}(t){\bf e}_{i}(0)=\tilde{U}^{\dag}(t)x_{i}(0)\tilde{U}{\bf
e}_{i}(\epsilon t).
\end{equation}
Using the relation (12) and the linear independence of ${\bf
e}_{i}$,we have
\begin{equation}\label{17}
x_{i}(t)=\tilde{U}^{\dag}(t)E_{ij}(\epsilon t)x _{j}(0)
\tilde{U}(t).
\end{equation}
However, $x_{i}(0)$ is a vector operator (first rank tensor), which means
\begin{equation}\label{18}
x_{i}(t)=\tilde{U}^{\dag}(t)R^{\dag}(\epsilon t)x
_{i}(0)R(\epsilon t) \tilde{U}(t),
\end{equation}
\begin{equation}\label{19}
R(\epsilon t) = P\exp\int\limits_{{\bf n}(0)}^{{\bf n}(\epsilon t)}\ ({\bf n}\times d{\bf n})\cdot{\bf
e}_{m}(0)\frac{-iL_{m}(0)}{\hbar},
\end{equation}
where $L_{m}(0) = ({\bf x}(0) \times {\bf P}(0)) \cdot{\bf e}_{m}(0)$.

In fact, the vector operator relation $E_{ij}(\epsilon t)x _{j}(0)=R^{\dag}(\epsilon t)x_{i}(0)R(\epsilon t)$ which
was used in proving (18) can be verified by comparing the derivatives of both sides with the aid of the commutation
relation $[x_{i}(0), P_{j}(0)] = i\hbar\delta_{ij}$. Similarly,

\begin{equation}\label{20}
P_{i}(t)=\tilde{U}^{\dag}(t)R^{\dag}(\epsilon t)P
_{i}(0)R(\epsilon t) \tilde{U}(t).
\end{equation}

Suppose $\tilde{U}(t)$ as appeared in equation (14) and (15) is known, then from (18) and (20), $U=R\tilde{U}$ is
the operator that solves the Heisenberg equations for $x_i$ and $P_i$. Therefore the task of finding $U(t)$ is now
reduced to the task of finding $\tilde{U}(t)$. Equations (17)-(20) embody the central point of our method, namely
the vector operator relations for $\bf x$ and $\bf P$ allow us to isolate the operator $R$ in the evolution operator
and to make full use of the vectors ${\bf e}_{1}$ and ${\bf e}_{2}$ in solving the Heisenberg equations so that the
other factor in the evolution operator can be easily found. In the following, $\tilde{U}(t)$ is determined, up to
$e^{i\alpha(t)}$, by analyzing the behavior of the solutions for $\tilde{x}_{i}(t)$ and $\tilde{P}_{i}(t)$.

The equations of motion (3) and (4) can be expressed in terms of the components $\tilde{x}_i$ and $\tilde{P}_i$. The
approach for finding these components described below is the same as that used in Ref \cite{qi} for solving a
related model. But the explanation here is self-contained. From equation (3), $\bf P$ is determined by knowing $\bf
x$. Substituting $\bf P$ from equation (3) into (4) and taking the dot product of the resulting equation with ${\bf
e}_i$, we obtain with the aid of (13) the following equations:

\begin{equation}
\ddot{\tilde{x}}_{1} - \omega_{c}\dot{\tilde{x}}_{2}= -
\frac{\omega_{c}}{2}\sigma_{2}\tilde{x}_{3}+
2\sigma_{1}\dot{\tilde{x}}_{3}+ \sigma_{1}^{2}\tilde{x}_{1}+
\sigma_{1}\sigma_{2}\tilde{x}_{2}+\dot{\sigma}_{1}\tilde{x}_{3},
\end{equation}
\begin{equation}
 \ddot{\tilde{x}}_{2} + \omega_{c}\dot{\tilde{x}}_{1} =
\frac{\omega_{c}}{2}\sigma_{1}\tilde{x}_{3} +
2\sigma_{2}\dot{\tilde{x}}_{3}+ \sigma_{2}^{2}\tilde{x}_{2}+
\sigma_{1}\sigma_{2}\tilde{x}_{1}+ \dot{\sigma}_{2}\tilde{x}_{3},
\end{equation}
\begin{equation}
\ddot{\tilde{x}}_{3} + \omega^{2}(\tilde{x}_{3} - a)=
\dot{e}_{3}^{2}\tilde{x}_{3}-
\frac{\omega_{c}}{2}\sigma_{2}\tilde{x}_{1}+
\frac{\omega_{c}}{2}\sigma_{1}\tilde{x}_{2}-
2\sigma_{1}\dot{\tilde{x}}_{1}- 2\sigma_{2}\dot{\tilde{x}}_{2}-
\dot{\sigma}_{1}\tilde{x}_{1} - \dot{\sigma}_{2}\tilde{x}_{2},
\end{equation}
where

\begin{equation}
 \sigma_{\mu}(\epsilon t) = \dot{\bf e}_{\mu}(\epsilon t) \cdot
 {\bf n}(\epsilon t);\,\, \mu=1,2.
\end{equation}
It is a useful fact that $\sigma_{\mu}(\epsilon t)$ is of the order of $\epsilon$ and $\dot{\sigma}_{\mu}$ is of the
order of $\epsilon^2$.

The many terms in (21), (22) and (23), though may seem complicated, all have simple physical origins. Take the term
$-\frac{\omega_{c}}{2}\sigma_{2}\tilde{x}_{3}$ on the right hand side of (21) for example. From equation (7), we
know that the frame ${{\bf e}_i}$ is rotating with the angular velocity ${\bf n}\times\dot{\bf n}$. Seen from the
stationary frame, $\tilde{x}_{3}$ actually moves with the velocity $\tilde{x}_{3}\dot{\bf n}$, which causes a
Lorentz force whose (acceleration) component along ${\bf e}_1$ is $-\omega_{c}\sigma_{2}\tilde{x}_{3}$. On the other
hand, the rotation of the magnetic field induces an electric field $-\frac{1}{c}\frac{\partial\bf A}{\partial t}$
whose acceleration along ${\bf e}_1$ is $\frac{\omega_{c}}{2}\sigma_{2}\tilde{x}_{3}$. The sum of these two gives
rise to the term in (21). Other terms which do not depend on $B$ on the right hand sides are due to inertial forces
associated with the rotation of the frame.

The method for treating the small terms on the right hand sides of (21)-(23) is to regard them as non-homogeneous
terms and to extract their contribution to the solution iteratively, as described below. Such an iteration procedure
is valid if it converges. Let us first consider the following simplified system:

\begin{equation}
\ddot{\tilde{x}}_{1} - \omega_{c}\dot{\tilde{x}}_{2}=
-\frac{\omega_{c}}{2}\sigma_{2}\tilde{x}_{3},
\end{equation}
\begin{equation}
\ddot{\tilde{x}}_{2} + \omega_{c}\dot{\tilde{x}}_{1} =
\frac{\omega_{c}}{2}\sigma_{1}\tilde{x}_{3},
\end{equation}
\begin{equation}
\ddot{\tilde{x}}_{3} + \omega^{2}(\tilde{x}_{3} - a) = 0.
\end{equation}
It amounts to neglecting most of the small terms on the right hand sides of (21)-(23). The solution to this system
is readily known because the terms $-\frac{\omega_{c}}{2}\sigma_{2}\tilde{x}_{3}$ and
$\frac{\omega_{c}}{2}\sigma_{1}\tilde{x}_{3}$ are non-homogeneous terms in (25) and (26) while $\tilde{x}_{3}$ is
known from (27). If the solution to the corresponding homogeneous system of (25) and (26) and the solution to (27)
are denoted as $\tilde{x}_{i}^{(0)}(t)$, then

\begin{equation}
\tilde{x}_{\mu}(t) = \tilde{x}_{\mu}^{(0)}(t) +
\frac{1}{2}\int\limits_{0}^t \sigma_{\mu}(\epsilon t')
\tilde{x}_{3}^{(0)}(t') dt',    (\mu=1, 2)
\end{equation}
while $\tilde{x}_{3}(t)=\tilde{x}_{3}^{(0)}(t)$.

The general iteration procedure for treating (21)-(23) is similar. By substituting $\tilde{x}_{i}^{(0)}(t)$ into the
right hand sides of (21)-(23), we will obtain the first order approximation to the solution and so forth. However,
the terms appeared in (21)-(23) but dropped in (25)-(27) do not contribute to the solution in the adiabatic limit of
$\mid\frac{\epsilon}{\omega-\omega_{c}}\mid\rightarrow0, \frac{\epsilon}{\omega}\rightarrow0$, and
$\frac{\epsilon}{\omega_c}\rightarrow0$. This is just the usefulness of the parallel transported vectors ${\bf
e}_{\mu}$, namely, the use of condition (13) has resulted in a simplification. So the system (21)-(23) and the
system (25)-(27) are equivalent in the adiabatic limit. The adiabatic limit defined above is reasonable and the case
of $\omega=\omega_{c}$ is excluded. Otherwise, there will be further degeneracy which is not of particular concern
to the magnetic field problem.

Therefore, in the adiabatic limit, (28) is the solution for $\tilde{x}_{\mu}(t)$ as appeared in (21)-(23). Now
$\tilde{x}_{3}^{(0)}(t) = a + [x_{3}(0)-a]\cos\omega t + \frac{\dot{x}_{3}(0)}{\omega}\sin\omega t$; the two
oscillating terms, when substituted into (28), vanish in the limit of $\epsilon \rightarrow 0$, therefore we have

\begin{equation}
\tilde{x}_{\mu}(t) =  \tilde{x}_{\mu}^{(0)}(t) - \frac{a}{2}
\int\limits_{{\bf n}(0)}^{{\bf n}(\epsilon t)}\ {\bf e}_{\mu}\cdot
d{\bf n},
\end{equation}
where use has been made of equation (24). In the above, the result of the integral is path-dependent, because ${\bf
e}_{\mu}$ is dependent on the path of $\bf n$. It is noticeable that although the shifting of the orbit as
represented by the integral is due to the electromagnetic force (as in equation (25) and (26)), the result is
independent on the magnitude of the magnetic field. By now, we have solved the Heisenberg equations (3) and (4) in
the adiabatic limit by making use of the two parallel transported unit vectors ${\bf e}_{\mu}$.

Now we turn to the problem of finding the operator $\tilde{U}$ in equation (14) and (15) that can realize (29).
Observe that although the integral in (29) is finite, its derivative with respect to time is of the order of
$\epsilon$ which goes to zero when $\epsilon \rightarrow 0$. Therefore, the velocity operator satisfies:
$\dot{\tilde{x}}_{i}(t) = \dot{\tilde{x}}_{i}^{(0)}(t)$. It follows from this that the Heisenberg picture
Hamiltonian (different from the Schr\"{o}dinger picture Hamiltonian because of its time dependence) of the system
(2) is an invariant in the adiabatic limit. Therefore equation (29) should be seen as embodying the adiabatic
theorem in a concrete way that is known only through solving the equations. Now it is important to observe that the
generator ${\bf P}(0)$ of the ordinary translation does not commute with the velocity operator. However, it is easy
to show that $[P_{i}(0)+\frac{q}{c}A_{i}(0), P_{j}(0)-\frac{q}{c}A_{j}(0)]=0$. Namely, a translation generated by $
{\bf P}(0)+\frac{q}{c}{\bf A}(0)$ can leave the velocity operator invariant. Such a translation is known as a
magnetic translation and is studied in Ref \cite{brown} and \cite{Zak}. The product of the dynamical operator
$e^{-\frac{i}{\hbar}H(0)t}$ which evolves $\tilde{x}_{i}(0)$ and $\tilde{P}_{i}(0)$ to $\tilde{x}_{i}^{(0)}(t)$ and
$\tilde{P}_{i}^{(0)}(t)$ respectively, and a magnetic translation operator which preserves
$\dot{\tilde{x}}_{i}^{(0)}(t)$ but shifts $\tilde{x}_{i}^{(0)}(t)$ to $\tilde{x}_{i}(t)$ qualifies as $\tilde{U}$:

\begin{equation}
\tilde{U}(t) = e ^{-\frac{i}{\hbar}H(0)t}M(\epsilon t),
\end{equation}
\begin{equation}
M(\epsilon t)= \exp\Bigg(-\frac{i}{\hbar}\Big(-\frac{a}{2}\int\limits_{{\bf n}(0)}^{{\bf n}(\epsilon t)}\ {\bf
e}_{\mu}\cdot d{\bf n}\Big){\bf e}_{\mu}(0)\cdot[{\bf P}(0)+ \frac{q}{c}{\bf A}(0)]\Bigg).
\end{equation}
With this, we determined the evolution operator up to a phase
factor in the following form:
\begin{equation}
  U(t) = R(\epsilon t)M(\epsilon t)D(t)
\end{equation}
where we used the fact that $M(\epsilon t)$ and $D(t)=e
^{-\frac{i}{\hbar}H(0)t}$ commute, and the operator $R$ is given
by equation (19).

The expression for $M$ as given by (31) can be verified directly
by checking equations (14) and (15). Intuitively, the magnetic
translation should be thought of as happening in the parallel
transported frame. Then through the operator $R$, we know how the
quantum system evolves in the frame specified by the Hamiltonian
(what we call the stationary frame). Notice that the displacement
vector in equation (31) is
\begin{equation}
{\bf d}(\epsilon t)=\Big(-\frac{a}{2}\int\limits_{{\bf n}(0)}^{{\bf n}(\epsilon t)}\ {\bf e}_{\mu}(\epsilon t')\cdot
d{\bf n}\Big){\bf e}_{\mu}(0),
\end{equation}
which agrees with the fact that inside the parallel transported frame, the coordinate axes should be seen as fixed,
i.e., ${\bf e}_{i}(0)$. By now, $U$ has been written as the product of a path-dependent geometrical operator and a
dynamical operator.

As mentioned in the introduction, the operator $U$ is not unique. Obviously, the exponential in (31) may be replaced
by the corresponding path-ordered exponential. It gives a different $U$ that gives the same $x_{i}(t)$ and
$P_{i}(t)$. The magnetic translation and its path-ordered exponential alternative are different because magnetic
translations along different directions do not commute. In the next section it is shown that it is the path-ordered
magnetic translation that enters into the evolution operator. Our purpose below is to find the relation between
magnetic translation and the corresponding path-ordered magnetic translation by using some simple properties of the
magnetic translation operator\cite{brown}.

Denote the path-ordered exponential as $M_{P}$. We have

\begin{equation}
M_{P}(\epsilon t)=e^{i\phi_{P}(\epsilon t)}M(\epsilon t).
\end{equation}
It has the advantage of easy differentiation, i.e.,
\begin{equation}
\frac{d}{dt}M_{P}(\epsilon t)=\Bigg(\frac{ia}{2\hbar}\big({\bf
e}_{\mu}\cdot\dot{\bf n}\big){\bf e}_{\mu}(0)\cdot[{\bf P}(0)+
\frac{q}{c}{\bf A}(0)]\Bigg)M_{P}(\epsilon t).
\end{equation}
To determine $\phi_{P}(\epsilon t)$ in equation (34), consider a
magnetic translation corresponding to the displacement $\bf d$,
\begin{equation}
M({\bf d})=\exp\big(-\frac{i}{\hbar}{\bf d}\cdot[{\bf P}(0)+ \frac{q}{c}{\bf A}(0)]\big).
\end{equation}
It is straight forward to verify that
\begin{equation}
M({\bf d}_{2})M({\bf d}_{1})=M({\bf d}_{1}+{\bf
d}_{2})exp\Big(-\frac{i}{2}({\bf d}_{1}\times{\bf
d}_{2})\cdot\frac{qB{\bf n}(0)}{\hbar c}\Big).
\end{equation}
This relation is the same as equation (9) in Ref \cite{brown}. (In \cite{brown}, $-e$ is the charge for the
electron, i.e., $q=-e$. Also, we use $M$ instead of $T$ for the magnetic translation operator since the latter is
reserved for the time of the adiabatic evolution.) Observe that $\frac{1}{2}({\bf d}_{1}\times{\bf d}_{2})\cdot
B{\bf n}(0)$ is the flux through the triangle formed by ${\bf d}_{1}$ and ${\bf d}_{2}$ with the tail of ${\bf
d}_{2}$ sitting on tip of ${\bf d}_{1}$. So it follows from equation (37) that for a sequence of magnetic
translations corresponding to displacement around a closed path $C$, $M_{P}=\exp\big(-\frac{iq}{\hbar
c}\Phi_{C}\big)$, where $\Phi_{C}$ is the magnetic flux through the loop $C$. (In \cite{brown}, the phase was
erroneously written as ``flux$(\frac{e}{2\hbar c})$", it should obviously be ``flux$(\frac{e}{\hbar c})".$) For an
open path, it follows from (37), the flux considered is through the loop formed by the curve ${\bf d}(\epsilon t')$
with $t'$ going from $0$ to $t$ and the straight line pointing from ${\bf d}(\epsilon t)$ to $\bf 0$. The area
enclosed by this loop is known because ${\bf d}$ is given by equation (33). The result for $\phi_{P}(\epsilon t)$
is:
\begin{equation}
\phi_{P}(\epsilon t)=-\Big(\frac{q}{\hbar
c}\Big)\frac{Ba^2}{4}\Big(\int\limits_{0}^t \sigma_{2}(\epsilon
t')dt'\int\limits_{0}^{t'} \sigma_{1}(\epsilon
t'')dt''-\frac{1}{2}\int\limits_{0}^t \sigma_{1}(\epsilon
t')dt'\int\limits_{0}^t \sigma_{2}(\epsilon t')dt'\Big).
\end{equation}
By the expression for $\sigma_{\mu}$ in equation (24),
$\phi_{P}(\epsilon t)$ is path-dependent just as expected.

\section{The Quantum Evolution Operator in a Factorised Form}
The operator $U$ obtained in the last section contains all the information from the Heisenberg equations. From the
fundamentals of quantum theory, we know that the Heisenberg picture and the Schr\"{o}dinger picture are equivalent.
If we consider a whole quantum system as governed by an adiabatically changing Hamiltonian, then an Abelian phase
factor is of course not physically observable \cite{solem}. However if we think of the adiabatic Hamiltonian as
governing one component of a quantum system that is later to interfere with the other component that has gone
through a different evolution, then the result is dependent on the phase content of each of the components. An
Abelian phase factor therefore is observable in an interference experiment. But it is important to bear in mind in
this case that the evolution of the other component is not governed by the adiabatic Hamiltonian. So far as a
complete quantum system is concerned, the Heisenberg picture and the Schr\"{o}dinger picture should contain equal
amounts of physical information. Because of the ambiguity in the choice of $U$ as mentioned at the end of the
introduction, it is impossible to determine completely the quantum evolution operator (and therefore the Berry phase
factor) through the Heisenberg picture.

The purpose of this section is to eliminate the ambiguity by making use of the Schr\"{o}dinger picture so that the
quantum evolution operator is determined completely. Let the evolution operator be $\cal U$. Then $\cal U$ is
related to $U$ by an Abelian phase factor,
\begin{equation}
{\cal
U}=e^{i\alpha}U=e^{i\alpha}RMD=e^{i\alpha}e^{-i\phi_{P}}RM_{P}D,
\end{equation}
where $\alpha$ is a real number. In the following, it is shown
that $\alpha=\phi_{P}(\epsilon t)$.

Consider an eigenstate $\mid\Psi(0)\big>$ of the initial Hamiltonian $H(0)$ that is a direct product of an
eigenstate of the harmonic oscillator along ${\bf n}(0)$ and a wave function describing motion perpendicular to
${\bf n}(0)$:
\begin{equation}
\mid\Psi(0)\big>=\mid\Psi_{\perp}(0)\big>\mid Osc(0)\big>.
\end{equation}
Such an eigenstate has the following property:
\begin{eqnarray}
      \big<\Psi(0)\mid\tilde{x}_3(t)\mid\Psi(0)\big>&=&a, \\
\big<\Psi(0)\mid\dot{\tilde{x}}_i(t)\mid\Psi(0)\big>&=&0.
\end{eqnarray}
All the $\mid\Psi(0)\big>$'s form a complete set of
eigenfunctions. At a later time, $\mid\Psi(0)\big>$ evolves into:
\begin{equation}
\mid\Psi(t)\big>={\cal U}(\epsilon
t)\mid\Psi(0)\big>=e^{i\alpha(\epsilon t)}R(\epsilon t)M(\epsilon
t)e^{-\frac{i}{\hbar}H(0)t}\mid\Psi(0)\big>.
\end{equation}
Since we already proved the adiabatic theorem in the previous section for the system (2), when substituting
$\mid\Psi(t)\big>$ into the Schr\"{o}dinger equation,

\begin{equation}
i\hbar\frac{d}{dt}\mid\Psi(t)\big>=H({\bf n}(\epsilon
t))\mid\Psi(t)\big>,
\end{equation}
we get the following condition:
\begin{equation}
\dot{\alpha}=i\big<\Psi(0)\mid
U^{\dag}(\dot{R}R^{\dag})U\mid\Psi(0)\big>+i\big<\Psi(0)\mid
\tilde{U}^{\dag}(\dot{M}M^{\dag})\tilde{U}\mid\Psi(0)\big>.
\end{equation}
The purpose is to determine $\alpha$. Observed that $\dot{\alpha}$ is necessarily of the order of $\epsilon$. But
upon integration ($t\in[0, 1/\epsilon]$) it has a finite result and indeed $\alpha$ is path-dependent. By equation
(19),
\begin{eqnarray}
iU^{\dag}\dot{R}R^{\dag}U&=&iU^{\dag}({\bf n}\times \dot{\bf
                            n})\cdot{\bf e}_{m}(0)\frac{-iL_{m}(0)}{\hbar}U,\nonumber\\
                        &=&U^{\dag}\frac{1}{\hbar}({\bf n}\times \dot{\bf n})\cdot{\bf
                            e}_{m}(\epsilon t) ({\bf x}(0) \times {\bf P}(0)) \cdot{\bf
                            e}_{m}(\epsilon t)U, \nonumber\\
                        &=&\frac{1}{\hbar}({\bf n}\times \dot{\bf n})\cdot{\bf
                               e}_{m}(\epsilon t) ({\bf x}(t) \times {\bf P}(t)) \cdot{\bf
                                e}_{m}(\epsilon t).
\end{eqnarray}
While according to equations (5) and (6), ${\bf x}(t) \times {\bf P}(t)=\tilde{x}_{i}(t)\tilde{P}_{j}(t){\bf
e}_{i}(\epsilon t)\times{\bf e}_{j}(\epsilon t).$ Since $\tilde{x}_{i}(t)$ and $\tilde{P}_{i}(t)$ are found in the
previous section in terms of $\tilde{x}_{i}^{(0)}(t)$ and the displacement vector, the calculation of
$i\big<U^{\dag}(\dot{R}R^{\dag})U\big>$ is reduced to the calculation of terms such as
$\big<\tilde{x}_{3}(t)\tilde{P}_{1}(t)\big>$, $\big< \tilde{P}_{3}(t)\tilde{x}_{1}(t)\big>$, etc., with coefficients
expressible in terms of $\sigma_{\mu}(\epsilon t)$ on account of equation (24). Observe from (45) and (46) that
$\dot{\alpha}$ is already of $\epsilon$ order due to $\dot{\bf n}$, so we do not need to keep $\epsilon$ order terms
from the products $\tilde{x}_{3}(t)\tilde{P}_{1}(t)$, $\tilde{P}_{3}(t)\tilde{x}_{1}(t)$, etc., because they make no
contribution to $\alpha$ in the adiabatic limit. The result is:
\begin{eqnarray}
i\big<\Psi(0)\mid U^{\dag}(\dot{R}R^{\dag})U\mid\Psi(0)\big>=\frac{qaB}{2c\hbar}\Big(\sigma_{1}\big<d_{2}+
\tilde{x}_{2}^{(0)}(t)\big>-\sigma_{2}\big<d_{1}+\tilde{x}_{2}^{(0)}(t)\big>\Big).
\end{eqnarray}
By making use of (34) and (35), the second term on the right hand side of (45) can also be calculated in a similar
way with the following result:
\begin{equation}
i\big<\Psi(0)\mid
\tilde{U}^{\dag}(\dot{M}M^{\dag})\tilde{U}\mid\Psi(0)\big>=\dot{\phi}_{P}-\frac{qaB}{2c\hbar}\Big(\sigma_{1}\big<d_{2}+
\tilde{x}_{2}^{(0)}(t)\big>-\sigma_{2}\big<d_{1}+\tilde{x}_{2}^{(0)}(t)\big>\Big).
\end{equation}
Therefore, with (45), (47), (48) and the initial condition $\alpha(0)=0$, we have:
\begin{equation}
\alpha(\epsilon t)=\phi_{P}(\epsilon t).
\end{equation}
Since $\alpha(\epsilon t)$ does not depend on the choice of $\mid\Psi(0)\big>$, it is the same for all initial
states. With this result and equation (39), the quantum evolution operator is determined to be:
\begin{equation}
{\cal U}(\epsilon t)=R(\epsilon t)M_{P}(\epsilon t)D(t),
\end{equation}
where $M_{P}(\epsilon t)$ is given by equations (34) and (38) in terms of the magnetic translation operator. Because
magnetic translations along two different directions do not commute, the Berry phase factor, if ${\bf n}(1)={\bf
n}(0)$, is essentially non-Abelian.

\section{Remarks and Possible Experiment}
In the following, some remarks are given. Experimental verification of the result (50) is also proposed.

(I) The infinite degeneracy requires solving the Heisenberg equations explicitly in order to take into account the
small perturbations in the rotating frame. Otherwise, say a confining potential is present in the two dimensional
plane and therefore eliminates the infinite degeneracy, the small perturbations would be averaged out. These are
quite different perturbations.

(II) Equation (29) also sheds light on the situation of no confinement in the direction of the magnetic field. For
such a case, motion along the direction of the magnetic field is not bounded. In the adiabatic limit (if such a
meaningful limit exists), $\tilde{x}_{3}\sim a$ would change greatly with time (because $t\in[0,T]$), leading to
complicated behavior in the $\tilde{x}_{1}\tilde{x}_{2}$ plane which may or may not obey any adiabatic theorem. At
least, none of the terms in the set of equations (21)-(23) can be neglected.

(III) The operator $R$ as given by formula (19) is determined (up to any time-dependent numerical phase factor) by
the tensor relations $E_{ij}(\epsilon t)x _{j}(0)=R^{\dag}(\epsilon t)x_{i}(0)R(\epsilon t)$ and $E_{ij}(\epsilon
t)P _{j}(0)=R^{\dag}(\epsilon t)P_{i}(0)R(\epsilon t)$. Notice especially that $-i$ in (19) cannot be replaced by
$i$; for a cyclic change (${\bf n}(1)={\bf n}(0)$) the latter results in ``$\exp(\frac{i}{\hbar}\Omega{\bf
L}(0)\cdot{\bf n}(0))$" while the correct result that follows from (19) is $R(1)$=$\exp(-\frac{i}{\hbar}\Omega{\bf
L}(0)\cdot{\bf n}(0))$, which is equivalent to the solid angle result well-known from the spin case \cite{berry}
except that due to the existence of magnetic translation, ${\bf L}(t)\cdot{\bf n}(\epsilon t)$ is not an adiabatic
invariant now. ($\Omega$ is the solid angle equal to the oriented area enclosed by the loop of $\bf n$ on the
two-sphere.)

(IV) The magnetic translation is a natural generalization \cite{brown}, \cite{Zak} of ordinary translation when a
magnetic field is present. The result (50) says that by rotating the magnetic field and the confining potential, a
wave packet will be displaced in the two dimensional plane perpendicular to the magnetic field direction in addition
to a rotation that follows that direction. The amount of displacement is predicted by (33). The transverse position
of the particle at the end of a cycle may be detected by withdrawing the confining potential and applying an
accelerating electric field along the magnetic field direction so that the particle can move out of the solenoid to
reach a detector.

I wish to thank Professor James M. Knight for beneficial contacts and many discussions. I also wish to thank
Professor Yakir Aharonov for a discussion and for suggesting the possibility of adding a potential along the
direction of the magnetic field. This work was presented in the Physics Department of the University of 
South Carolina on
Nov 22, 1999.

\end{document}